\documentclass[aps,prl,twocolumn,showpacs,amsmath,amssymb,superscriptaddress]{revtex4-2}
\usepackage{graphicx}
\usepackage{CJK}
\usepackage{color}
\usepackage{bm}
\usepackage[colorlinks,bookmarks=false,citecolor=blue,linkcolor=red,urlcolor=blue]{hyperref}
\usepackage{multirow}
\usepackage{ulem}
\usepackage{tabularx}
\usepackage[nounderscore]{syntax}
\graphicspath{{fig/}}

\newcommand{\be}{\begin{equation}}
\newcommand{\ee}{\end{equation}}

\newcommand{\trace}{{\rm Tr}}

\begin{document}
%\begin{CJK*}{UTF8}{gbsn}
\title{Coplanar order induced by emergent frustration}
\author{Zehui Deng}
\altaffiliation{The two authors contributed equally to this work.}
\affiliation{School of Physics and Electronics, Hunan Normal University, Changsha 410081, China}
\affiliation{Beijing Computational Science Research Center, Beijing 100193, China}
\affiliation{Key Laboratory of Multiscale Spin Physics (Ministry of Education), Beijing Normal University, Beijing 100875, China}
\author{Lu Liu}
\altaffiliation{The two authors contributed equally to this work.}
\affiliation{School of Physics, Beijing Institute of Technology, Beijing 100081, China}
\author{Wenan Guo}
\email{waguo@bnu.edu.cn}
 \affiliation{School of Physics and Astronomy, Beijing Normal University, Beijing 100875, China}
 \affiliation{Key Laboratory of Multiscale Spin Physics (Ministry of Education), Beijing Normal University, Beijing 100875, China}
\author{Hai-Qing Lin}
\email{hqlin@zju.edu.cn}
%\affiliation{Beijing Computational Science Research Center, Beijing 100193, China}
%\affiliation{Department of Physics, Zhejiang University, Hangzhou 310027, China}
\affiliation{Institute for Advanced Study in Physics and School of Physics, Zhejiang University, Hangzhou 310058, China}
\date{\today}

\begin{abstract}
Traditional frustration arises from the conflict between the spin alignments due to the geometry or the nature of the interactions. Here, we demonstrate a novel form of frustration, dubbed ``emergent frustration'', 
which is induced by the symmetry that emerges at the phase transition point of a quantum spin model devoid of geometric frustration.
We study the two-dimensional bipartite chequerboard $J$-$Q$ model, which hosts the antiferromagnetic (AFM) state to the plaquette-singlet solid state (PSS) phase transition detected in the Shastry-Sutherland compound SrCu$_2({\rm BO}_3)_2$. 
By analyzing the scaling behavior of the R\'enyi entanglement entropy with smooth 
boundaries at the transition point, we observe an unexpected scaling behavior, which indicates that the number of Goldstone modes is five. 
We explain this by proposing a novel scenario in which the system is described by an effective quantum rotor Hamiltonian with a three-sublattice geometry that frustrates collinear order while supporting coplanar order.
Such a three-sublattice geometry arises from the emergent symmetry of coexisting 
orders, which 
may also occur at the AFM-PSS transition point of SrCu$_2({\rm BO}_3)_2$. Therefore, experimental investigations are warranted. 
\end{abstract}

\maketitle

%\section{Introduction}
\textit{\color{blue}Introduction.---}
Frustrated quantum antiferromagnets have been at the center of intense experimental and theoretical investigations for decades\cite{Starykh_2015, SCHMIDT2017}.
One reason for this is the search for quantum spin liquid states,  which are related to high-temperature superconductivity. 
Another reason is that frustrated magnetic materials also feature unusual ordered states, e.g., the non-collinear order always arises as a consequence of frustration, which is typically 
geometric\cite{Olsen2024AntiferromagnetismIT}. Antiferromagnets 
in the triangular lattice and the Kagome lattice, which are non-bipartite and have three magnetic sublattices, are the two prominent geometric-frustrated examples \cite{Tchernyshyov}.
Even when the influence of quantum fluctuation is included, it has been confirmed that
the quantum spin-1/2 model remains ordered in the classical $120^\circ$ pattern\cite{Huse-PRL1988, Bernu-PRL1992, WhitePRL1999}.
In this work, we demonstrate an unusual {\it emergent frustration} induced by an emergent symmetry at the phase transition point of an antiferromagnetic quantum spin system, the chequerboard $J$-$Q$ (CBJQ) model, defined on a bipartite lattice. 
As a result, an unconventional coplanar order breaking the emergent $O(4)$ symmetry presents at the phase transition point. 

The chequerboard $J$-$Q$ model is introduced \cite{zhaobw-np} as a 
designer Hamiltonian \cite{Kaul-bridge} amenable to large-scale quantum Monte Carlo (QMC) simulations to mimic
the quantum phase transition from an antiferromagnetic(AFM) state to a plaquette-singlet 
solid (PSS) in the quasi-2D Shastry-Sutherland (SS) compound SrCu$_2({\rm BO}_3)_2$ \cite{ZayedPSS2017, MilaNature2021, LLSunPRL2020, Liulu-Science, guo2025deconfined}. The transition is widely regarded as the most promising candidate for the deconfined quantum critical point (DQCP),
a direct continuous phase transition between two unrelated ordered states, beyond the paradigm of Landau-Ginzburg-Wilson\cite{Senthil, SenthilPRB}. The nature 
of the transition is still under debate \cite{corbozpRB2013, LiuWYprl2024, WanglingSS2024, LeeJYPRX2019, zhaoerhai2022, XiNingPRB2023}.
Although the CBJQ and SS models are different at the lattice level, 
one expects them to describe the same universal physics of the AFM-PSS transition according to the symmetries involved, based on the renormalization group theory. Indeed, it is demonstrated 
in \cite{LLSunPRL2020}
that the CBJQ model correctly depicts the thermodynamic behavior of the SrCu$_2({\rm BO}_3)_2$, 
where a first-order phase transition occurs between AFM and PSS\cite{zhaobw-np}. The 
bi-critical behavior of the SrCu$_2({\rm BO}_3)_2$ at finite temperature is also predicted by the CBJQ model and was later obtained in experiment\cite{guo2025deconfined}. 
Additionally, the CBJQ model successfully describes the AFM-PSS transition in 
SrCu$_2({\rm BO}_3)_2$ when a magnetic field is applied \cite{Liulu-Science}.

Here, we study the AFM-PSS phase transition of the CBJQ model by analyzing the scaling behavior of the R\'enyi entanglement entropy (EE),
which is a valuable probe of quantum many-body systems\cite{LAFLORENCIE2016}.
Through generalizing the scaling formula of EE to a coplanar ordered 
system that breaks the $O(n)$ continuous symmetry, we demonstrate that the scaling behavior 
observed using QMC simulations reveals an unexpected three-sublattice coplanar order breaking an emergent $O(4)$ symmetry on the bipartite lattice.  
We propose a novel scenario in which the system is described by an effective quantum rotor Hamiltonian with a three-sublattice geometry that frustrates collinear order but supports coplanar order. 
Such a three-sublattice geometry arises from the emergent symmetry of coexisting orders, and the resulting coplanar order 
may also present at the AFM-PSS transition point of SrCu$_2({\rm BO}_3)_2$. Therefore, experimental investigations are warranted.

\begin{figure}[h]
     \centering
     \includegraphics[width=0.45\textwidth]{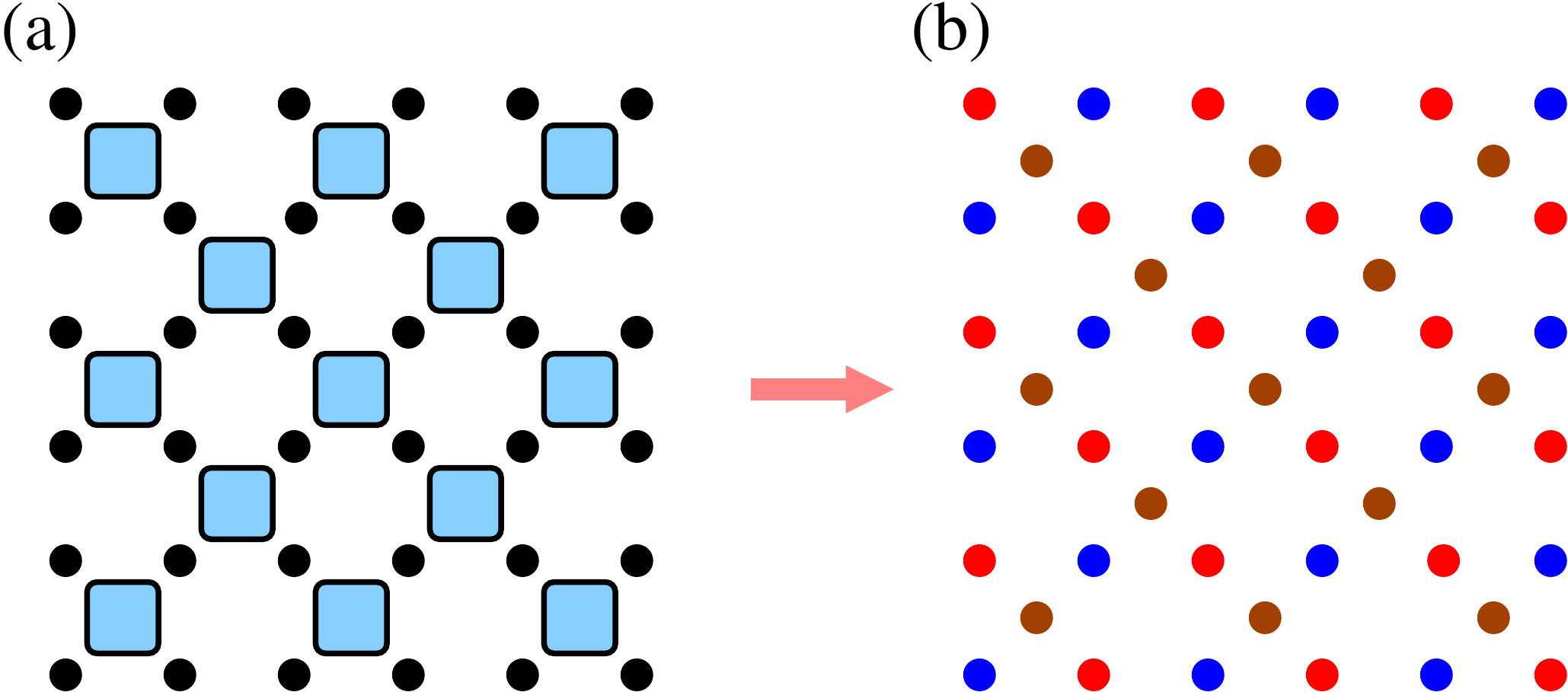}
   \caption{(a) The CBJQ quantum spin model and (b) its effective super spin model at the AFM-PSS transition point. In the CBJQ model, the black dots represent lattice sites. The light blue plaquettes represent the four-spin $Q$ terms.
   In the superspin model, the red, blue, and brown sites represent three sublattices,
respectively. The four-component super spins sitting on the lattice form a coplanar order breaking the emergent $O(4)$ symmetry.
   }
      \label{model1} 
 \end{figure}

\textit{\color{blue}Model and emergent frustration.---}
The Hamiltonian of the CBJQ model is defined using the singlet projection operator $P_{ij}=(1/4-{\bm S}_{i}\cdot {\bm S}_{j})$,
\begin{equation}
    H=-J\sum_{\langle ij \rangle}P_{ij}-Q\sum_{ijkl\in \square^{\prime}}(P_{ij}P_{kl}+P_{ik}P_{jl}),
    \label{model}
\end{equation}
where site pairs $\langle ij \rangle $ are nearest neighbors on a periodic square lattice with 
$N=L^2$ sites and $\square^{\prime}$ denotes the $2\times2$ $Q$-plaquettes, see 
Fig.\ref{model1}(a). We set $J=1$ to fix the energy scale.  

The model hosts a single quantum phase transition at $Q_c=4.5977(1)$ from the PSS state to the AFM 
state\cite{zhaobw-np}.
The PSS and AFM order parameters can be defined as 
\begin{equation}
     m_p=\frac{2}{N}\sum_{\bm c}m_p({\bm c}), ~~~~{\bm m}=\frac{1}{N}\sum_{\bm r} {\bm m}({\bm r}),
\label{psorder}
\end{equation}
respectively.
The local PSS order parameter $m_p({\bm c})=\theta({\bm c}) \Pi({\bm c})$, $\Pi({\bm c})=S^z({\bm c}_1)S^z({\bm c}_2)S^z({\bm c}_3)S^z({\bm c}_4)$, where the index ${\bm c}$ runs over the \textit{center} of $Q$ plaquettes in Fig.~\ref{model1}(a)
and ${\bm c}_i (i=1,2,3,4)$ labels the four corner sites of the plaquette; $\theta({\bm c})=\pm 1$ corresponds to even or odd plaquette rows, respectively. 
The local AFM order parameter ${\bm m}({\bm r})=\theta({\bm r}) {\bm S}({\bm r})$, ${\bm r}$ runs over $N$ lattice sites and $\theta({\bm r})=\pm 1$ is the staggered AFM sign.

The more conventional 2D square lattice $J$-$Q$ models \cite{Sandvik2007} have a columnar valence-bond solid (CVBS) phase, besides the AFM phase.
The CVBS order parameter is defined by a two-component vector ${\bm D}=(D_x,D_y)$, with
\begin{equation}
        D_\alpha = \frac{1}{N}\sum_{\bm r} D_\alpha({\bm r}),
        \label{dimerorder}
\end{equation}
where the local order parameter $ D_\alpha({\bm r})=(-1)^{{\bm r}_\alpha}
{\bm S}({\bm r}) \cdot {\bm S}({\bm r}+\hat{\alpha}),$ where $\alpha=x, y$, $\hat{\alpha}$ is the unit vector in $\alpha$ direction and ${\bm r}$ runs over all lattice sites.
The AFM-CVBS transition of the model can be described by a field theory with topological 
Wess-Zumino-Witten term\cite{Senthil-NLSM},
where a five-component order parameter is formed by combining the AFM and CVBS order parameters ${\bm \phi}=({\bm m}, D_x, D_y)$. 
The corresponding superspins ${\bm S}^{(\rm s)}({\bm r})$ can be defined 
with five components, $ S^{(\rm s)}_{\alpha}({\bm r}), (\alpha=1,\cdots,5) $, where $S^{(\rm s)}_{1,2,3}({\bm r}) = m_{x,y,z}({\bm r})$,
and $S^{(\rm s)}_{4,5}({\bm r})= D_{x,y}({\bm r})$, respectively.
%in Eq. (\ref{dimerorder}), 
Since ${\bm r}$ in both ${\bm m}$ and ${\bm D}$ runs over lattice sites, this allows the superspins ${\bm S}^{(\rm s)}({\bm r})$ locate at lattice site ${\bm r}$.
At the transition point, the model has an emergent $O(5)$ symmetry\cite{Nahum-SO5}.
%Recent work \cite{deng2024} has shown that 
It is further demonstrated that the transition is first order with the emergent $O(5)$ symmetry broken down to $O(4)$ symmetry featuring a collinear order\cite{deng2024}. This 
can be understood by assuming the emergent symmetry inherited from a neighboring multicritical point hosting emergent $SO(5)$ symmetry separating a DQCP line and the first-order transition line \cite{Takahashi-multicri,chester2024}, or that the $SO(5)$ symmetry of a nearby nonunitary 
fixed point survives to length scales larger than the correlation length in the complex conformal field theory for the pseudocritical behavior observed for the model\cite{Nahum-PRX2015, WangchongPRX2017}. 

For the current CBJQ model, the AFM-PSS transition is first order, with the symmetry enhanced to $O(4)$\cite{zhaobw-np}. 
It is then tempting to construct a similar field theory describing the AFM-PSS 
transition based on a four-component order parameter formed with the AFM and PSS 
order parameters ${\bm \phi}=({\bm m}, m_p)$. The corresponding superspins satisfy
$ S^{(\rm s)}_{1,2,3}({\bm r'}) = m_{x,y,z}({\bm r'})$, % for $\alpha=1,2,3$,  
and $ S^{(\rm s)}_{4}({\bm r'})= m_{p}({\bm r'})$.
However, at the microscopic lattice level, there is a big difference from the conventional $J$-$Q$ model: 
the local PSS order parameter $m_p({\bm r'})$ is defined at the center of $Q$-plaquettes,  therefore, the locations of the superspins,  ${\bm r'}$, are not restricted to the sites of the 
square lattice but can also be positioned at the centers of the $Q$-plaquettes, forming a kagome-like 
lattice with three sublattices, as illustrated in Fig.~\ref{model1}(b).

Assume the system is described by an effective Hamiltonian of a quantum rotor 
\be
H=\frac{{\bm S}^{(\rm s)}\cdot {\bm S}^{(\rm s)}-({\bm S}^{(\rm s)}_A)^2-({\bm S}^{(\rm s)}_B)^2-({\bm S}^{(\rm s)}_C)^2}{2 I L^2},
\label{O4rotor1}
\ee
with ${\bm S}^{(\rm s)}$ (${\bm S}^{(\rm s)}_{A,B,C}$) the total superspin of the system(three sublattices), $I$ the inertia moment density, and $L$ the linear size of the system. We expect naturally a coplanar order\cite{Bernu-PRL1992, Bernu_triangularHeis, Lavalle_spectrum, Azaria, Kolley}, as what happens for the SU(2) AFM Heisenberg model on the frustrated three-sublattice systems. 
The degeneracy of energy level with total angular quantum number $S$ is then $(S+1)^4$, which is
the special case of the degeneracy $\sim S^{(2n-4)}$ for a coplanar ordered system with $O(n)$ symmetry for $n=4$.
We will demonstrate below through scaling analysis of EE that this is indeed the case.  

\textit{\color{blue}Scaling of EE for systems with coplanar order.---}
Inspired by numerical results \cite{Kallin-EE,ee-spinwave2011}, Metlitski and Grover predict that
a logarithmic term with a coefficient proportional to the number of Goldstone modes $N_G=n-1$
is present in the scaling of EE with smooth boundaries for a collinear antiferromagnet with continuous symmetry broken from $O(n)$ to $O(n-1)$, in addition to the area law %when the boundary is smooth
\cite{Metlitski-EE},
\begin{equation}
	S_\alpha(L) =a L^{d-1}+b \ln(\frac{\rho_{s}}{c}L^{d-1})+\gamma_{\rm ord},
    \label{fssS2}
\end{equation}
where $L$ is the system size, $d$ is the spatial dimension,  
$\rho_s$ is the spin stiffness and $c$ is the spinwave velocity. $\gamma_{\rm ord}$ is a universal geometry-dependent finite constant, as all the 
short-distance physics are absorbed into $\rho_s$ and $c$. The coefficient $b=N_G/2$, with $N_G$ the 
number of Goldstone modes of the collinear ordered state.
It has been shown that this formula only works for very large system sizes\cite{D'Emidio, Helmes-Wessel, zhao2022measuring}, very 
strong order\cite{D'Emidio}, or $O(2)$ continuous symmetry\cite{Kulchytskyy}. 
An improved scaling formula \cite{Deng_PRB} is later derived as follows,
 \begin{equation}
        S_\alpha(L) =a L^{d-1} + b \ln(I(L)^{1/2}\rho_s(L)^{1/2} L^{d-1}) +\gamma_{\rm ord},
        \label{newS2}
\end{equation}where $I(L)$ is the finite-size inertia moment density of the quantum rotor model, which describes the energy spectrum of the $O(n)$ ordered phase, and $b=N_G/2$.  

Metlitski and Grover's formula, Eq.(\ref{fssS2}), is derived under the assumption 
that the order is collinear. 
Rademaker\cite{Rademaker} extends the formula to the coplanar ordered antiferromagnet, breaking the $O(3)$ symmetry with $N_G=3$. 
For more general cases, the coplanar ordered antiferromagnet with $O(n)$ symmetry broken, Eq.(\ref{fssS2}), and thus Eq.(\ref{newS2}), should also 
apply, but with $N_G$ counting the Goldstone modes of the coplanar order, which is $N_G=2n-3$.
For details of a derivation, see supplemental material %~\footnote{See Supplemental Material for a derivation of the EE scaling in $O(n)$ coplanar ordered system, polynomial fit of spin stiffness and susceptibility.}
\cite{supplemental}. 

\textit{\color{blue}Numerical results and scaling analysis.---}
The R\'enyi EE is defined as
\begin{equation}
    S_{\alpha}(A)=\frac{1}{1-\alpha}\ln{\trace [\rho_{A}^{\alpha}]},
    \label{Sn_def}
\end{equation}
where $\alpha$ is the R\'enyi index ($\alpha=2$ in our work), and 
$\rho_{A}=\trace_{\bar{A}}{\rho}$ is the reduced density matrix of a 
subsystem $A$, with $\bar{A}$ its complement and $\rho$ the 
density operator. 
In the QMC simulations, we consider an $L\times L$ square lattice with periodic boundary conditions 
employed in both lattice directions. In particular, we bipartite the toroidal lattice into two equally sized cylindrical strips of size 
$N_{A}=L/2\times L=N_{\bar{A}}$ containing no corners and study the R\'enyi EE of one subregion.

\begin{figure}[h]
    \centering
    \includegraphics[width=0.45\textwidth]{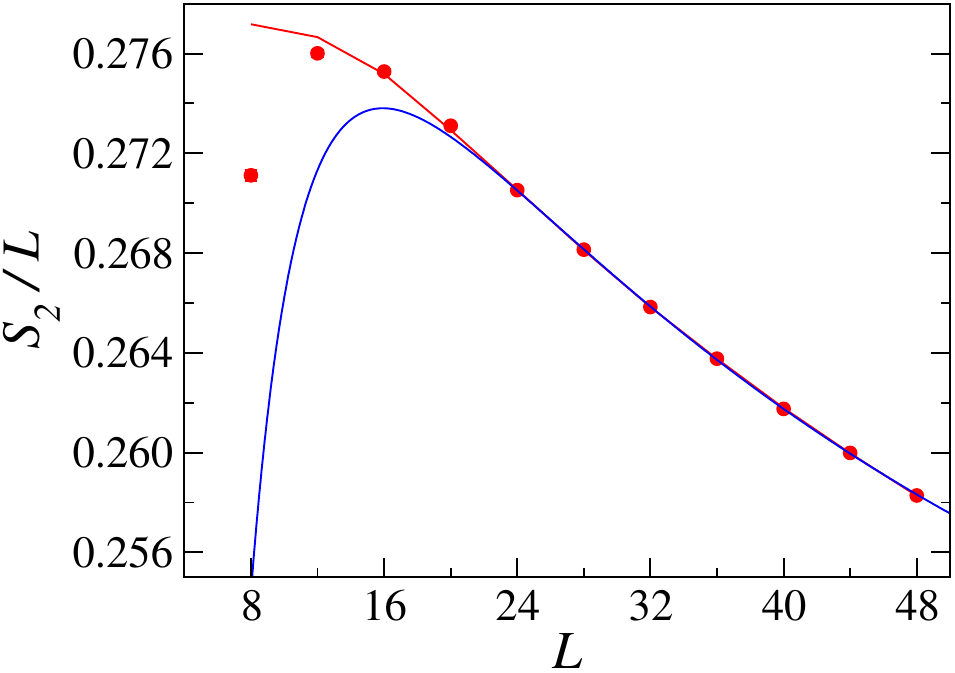}
    \caption{$S_2(L)/L$ (red dots) versus system size $L$ at $Q_c=4.5977$ of the CBJQ model.  
    The red solid line is the fit of Eq. (\ref{newS2}) with $I(L)$ obtained from chiral perturbation theory for the EE data of sizes $L\geq L_{\rm min}=24$ and $N_G/2$ is found to be $2.46(9)$, see Tab. {\ref{chiralperturbation-tab:fitlog_mod1}} for details. 
  The solid blue line shows the general fit function $S_2(L)=aL+b\ln(L)+c$ for data of sizes $L\geq L_{\rm min}=24$.
  }
    \label{fig:S2-L}
\end{figure}

With the help of the replica trick\cite{Calabrese2004}, 
$S_\alpha(A)$ can be expressed as the ratio of free energies, which can be calculated by  
using the nonequilibrium work algorithm developed recently\cite{Alba-entropy, D'Emidio}.
In this work, we make use of the version for the projector QMC method \cite{sandvik-pqmc, pqmc-loopupdate} to extract $S_2(A)$\cite{D'Emidio}.
To guarantee the accuracy of the EE, we compute 2000 to 3000 nonequilibrium 
work realizations for each system size ranging from $L=8$ to $48$. Each work realization consists of 
$N_{A}\times 10,000$ nonequilibrium time steps. The projection power is set to $m=2L^3$, which is large enough to probe the ground state properties of the models under investigation. The obtained $S_2(L)$ versus system size $L$ at $Q_c$ of the CBJQ model is shown in Fig. \ref{fig:S2-L}.

We now study the scaling behavior of $S_2(L)$ at $Q_c$.  

First, we try to fit Eq. (\ref{fssS2}) to the data. 
Table \ref{tab:fitlog} shows details of the fitting procedure. 
The formula can be fitted for all data sets with $L_{\rm min} \ge 20$. The curve of the fitting function is shown in Fig. \ref{fig:S2-L}.
However, the value of $b$ obtained is far from $N_G/2=1$ for the AFM ordered state, nor is it close to $N_G/2=1.5$ 
for the collinear order with symmetry breaking from $O(4)$ to $O(3)$. 
Moreover, the fitted parameters drift as small-size points are excluded gradually,
suggesting the fits are not stable and systematic errors exist.

This phenomenon aligns with prior observations 
that Eq.~(\ref{fssS2}) only works for very large systems, or the order is strongly 
enhanced. To capture the anticipated logarithmic correction due to spontaneous breaking of continuous symmetry, we have to 
utilize the improved scaling formula Eq. (\ref{newS2}) in the fitting of $S_2(L)$ data. 
To do so, we need finite-size values of the spin stiffness and the inertia moment density as inputs. 
\begin{table}[thb]
  \caption{ Fitting results of Eq. (\ref{fssS2}) to $S_2(L)$ at $Q_c$ with $L=16-48$. 
	Here $\gamma'=b \ln (\rho_s/c)+\gamma_{\rm ord}$.	
	}
    \begin{tabular}{c|c|c|c|c}
    \hline
     \hline
        $L_{\rm min}$ &  $a$         & $b$        & $\gamma'$                       &$\chi_r^2$/P-value \\   
    \hline
                16 &  0.2257(4)   & 0.71(1)   &  -1.17(3)                & 3.33/0.003   \\
    \hline
               20 &  0.2245(5)   & 0.75(2)   &  -1.27(4)               & 2.10/0.06\\
    \hline
               24 &  0.2226(9)   & 0.82(3)   &  -1.44(7)               & 0.53/0.72 \\
    \hline
               28&  0.222(2)   & 0.85(6)  &  -1.5(2)               & 0.51/0.68\\     
     \hline
    \hline
     \end{tabular}
     \label{tab:fitlog}
 \end{table}

 \begin{figure}[h]
     \centering
     \includegraphics[width=0.45\textwidth]{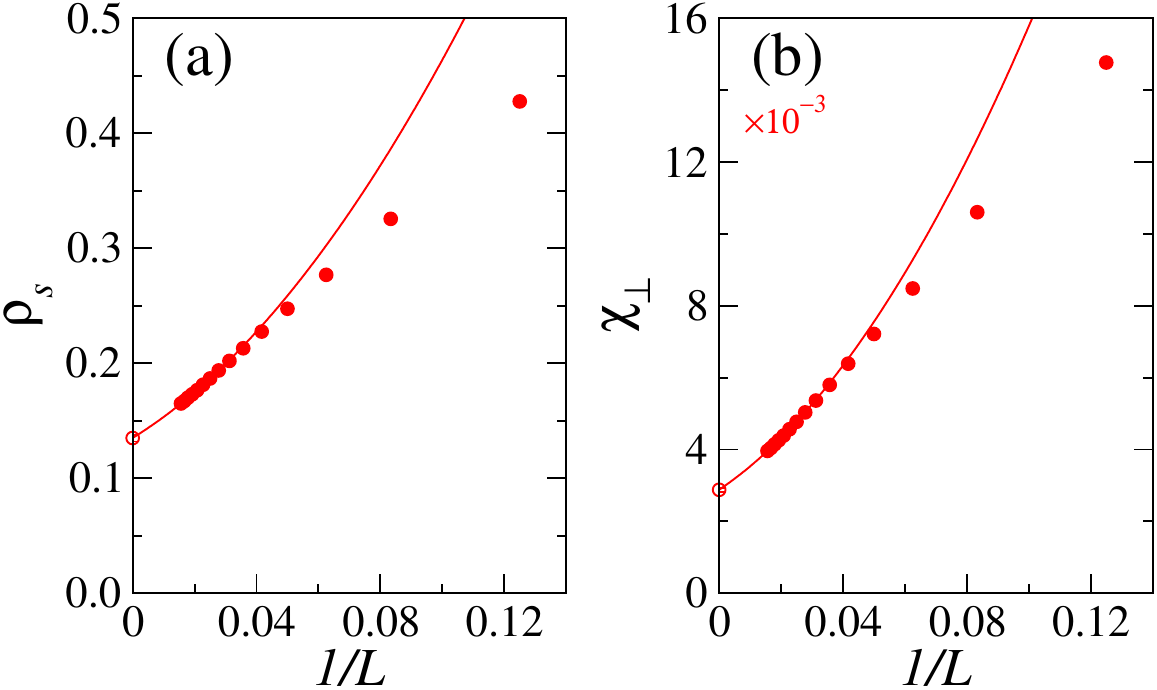}
	 \caption{ Spin stiffness $\rho_s$ (a) and transverse susceptibility $\chi_\perp$ (b)
     versus $1/L$ at $Q_c$. The solid red lines are polynomial fits to data with $L\ge 36$. The open circles show $\rho_s(\infty)=0.135(3)$ and $\chi_\perp=0.00288(6)$, respectively. 
     }
     \label{rhosqc}
 \end{figure}

We calculate the spin stiffness $\rho_s(L)$ using the stochastic series expansion
(SSE) QMC method with loop updates at inverse temperature $\beta=2L$ \cite{Sandvik1997, Sandvik-review}
through the fluctuations of the winding number $W_\alpha$ of spin transporting
for a finite lattice, where the symmetry is not broken. A factor $n$ is included to account for $O(n)$ rotational symmetry. 
Here, we have set $n$ to be the tempted value 4. The choice of $n$ could change neither the presence of the logarithmic term nor the coefficient of the 
term, except for $\gamma_{\rm ord}$; thus, it will not bias our conclusion. The correctness will be shown self-consistently. The obtained $\rho_s(L)$ are plotted in Fig. \ref{rhosqc}(a).

The inertia moment density $I(L)$ is related to the energy of the tower of excited states of the $O(n)$ quantum rotor described by Eq.(\ref{O4rotor1}), which characterizes the CBJQ model at the AFM-PSS transition point, where $S$ denotes the total superspin of the system: 
\begin{equation}
	E_L(S)=\frac{S (S+n-2)}{2L^2 I(L)}.
 \label{energylevels1}
\end{equation}
Here, we have set $E_L(0)=0$ and assume $n=4$. 
At the thermodynamic limit, $I(L)$ converges to the transverse susceptibility $\chi_\perp$
in the way predicted by the 
chiral perturbation theory \cite{chiral-perturbation} 
up to $1/L$ \cite{Neuberger_fss_AFH, FisherDS_fss_AF, Lavalle_spectrum}: 
 \begin{equation}
	I(L)=\chi_\perp[1+\frac{(n-2)}{c\chi_\perp L}\frac{3.900265}{4\pi}+O(\frac{1}{L^2})].
   \label{inertia}
\end{equation}
%Here, $\chi_\perp$ is the transverse susceptibility in the thermodynamic limit, which 
$I(L)$ can then be obtained through $\chi_\perp$, which is related to the uniform susceptibility $\chi_u$ defined by the fluctuation of total magnetization at zero temperature through $\chi_\perp=2\chi_u$. 
%for the system with $O(4)$ symmetry. 

However, for a finite system, $\chi_u$  vanishes at $T\to 0$, since the total magnetization is conserved, and the ground state is a singlet. 
We calculate 
the wave-vector ${\bm q}$-dependent susceptibility $\chi({\bm q})$, which is the Fourier transform 
of the static spin-spin susceptibility 
in real space $\chi(k,l)$ given by the Kubo integral,
which is obtained using standard SSE simulations\cite{Sandvik1997}. Here, $k,l$ denote the spin.
The value of $\chi({\bm q})$ at the longest wavelength, ${\bm q}=(2\pi/L,0)$, is taken as the definition of the 
finite-size uniform susceptibility $\chi_u(L)$,
%=\chi(2\pi/L,0)$, 
which converges to $\chi_u$ when $L \to \infty$\cite{Wang-dblh}.
The obtained $\chi_\perp(L)=2\chi_u(L)$ are shown in Fig. \ref{rhosqc}(b). 

%We analyze $\chi_{\perp}(L)$ and $\rho_{s}(L)$ to obtain 
The thermodynamic limit value $\chi_\perp$ and $\rho_{s}$ are obtained using polynomial fitting. For details, see \cite{supplemental}.
Our final estimates are $\chi_{\perp}=0.00288(6)$ and $\rho_{s}=0.135(3)$,  which leads to $c=\sqrt{\rho_s/\chi_\perp}=6.846.$ 
With $n=4$ for the CBJQ model at the transition point, and substituting $c$ and $\chi_{\perp}$ into (\ref{inertia}), we have %the specific form for the  inertia moment density,
 \begin{equation}
    I(L)=0.00288\times[1+\frac{31.5}{L}+O(\frac{1}{L^2})].
    \label{inertia1}
\end{equation}

 \begin{table}[thb]
	 \caption{ Fitting results of Eq.(\ref{newS2}) to $S_2(L)$ at $Q_c$ with $L=8-48$.
	  $I(L)$ obtained from chiral perturbation theory and $\rho_s(L)$ are used as inputs.}
    \begin{tabular}{c|c|c|c|c}
    \hline
     \hline
         $L_{\rm min}$ &  a           & $b$    &$\gamma_{\rm ord}$ &$\chi_r^2$/P-value \\   
    \hline
                12 &  0.2037(5)   & 2.53(3)   & 1.78(2)                 & 3.00/0.004   \\
    \hline
                16 &  0.2055(7)   & 2.43(4)    & 1.73(2)                  &0.66/0.68 \\
   \hline
                20 &  0.206(1)   & 2.38(6)    & 1.70(3)                  & 0.42/0.83 \\
      \hline
                24&  0.205(2)   & 2.46(9)    & 1.74(4)                   & 0.18/0.95\\     
     \hline
                28&  0.206(3)   & 2.4(2)    & 1.71(7)                   & 0.13/0.94\\  
             
    \hline
    \hline
     \end{tabular}
     \label{chiralperturbation-tab:fitlog_mod1}     
 \end{table}

Fitting results of Eq.(\ref{newS2}) to $S_2(L)$ using $I(L)$ obtained in Eq. (\ref{inertia1}) and $\rho_s(L)$ as inputs
are listed in Tab. \ref{chiralperturbation-tab:fitlog_mod1}.
The results of $b$ approach 2.5 within about one error bar for $L_{\rm min}=16$. 
The fits remain stable upon further excluding small-sized points by gradually increasing $L_{\rm min}$, although error bars on the fitting parameters grow
rapidly.

Despite inherent systematical errors due to numerical errors in $I(L)$ and $\rho_s(L)$ and 
ignoring higher-order corrections, our fitting procedure indicates that systematic errors in the fits have been 
removed by excluding data of sizes smaller than $L_{\rm min}$, akin to the fitting in the 2D AF Heisenberg and 
bilayer Heisenberg models\cite{Deng_PRB}.
Hence, we may safely assert that the system is ordered with $b=2.5$ from fits with $L_{\rm min} \ge 24$. This result suggests that $N_G=5$
for a coplanar ordering of an $O(4)$-symmetric
system with three sublattices, which frustrates collinear order.

\textit{\color{blue}Discussion and Conclusion.---}
%\textit{\color{blue} Coplanar order scenario.---}
We have analyzed the scaling behavior of the R\'enyi EE with smooth boundaries at the AFM-PSS 
transition point of the CBJQ model and identified an unexpected novel scaling behavior.
Earlier research proposed \cite{zhaobw-np}, in the transition region, the system is described by an effective quantum XXXZ model,
where the ratio $Q/J$ tunes the order parameter from the AFM state through the emergent $O(4)$ point into the PSS state.
Naively, one expects that, at the thermodynamic limit, the $O(4)$ 
symmetry breaks into its $O(3)$ subgroup forming a collinear order, 
leading to $N_G=3$ and $b=1.5$ in Eq.(\ref{newS2}).  
Our result $N_G=5$ above negates such a view, suggesting a coplanar order breaking the emergent $O(4)$ symmetry. This finding can be explained by the scenario discussed in the previous part of this paper, that the superspins are positioned on a kagome-like lattice, described by an effective
quantum rotor Hamiltonian, in which the {\it emergent} three-sublattice geometry
frustrates the collinear order, but supports the coplanar order. 
Such a novel frustration and the induced coplanar order may appear in the AFM-PSS transition of the SrCu$_2({\rm BO}_3)_2$. Experimental investigations are called for.

\begin{acknowledgments} 
\textit{\color{blue}Acknowledgments.---}
We would like to thank Bin Zhou and Yuan Li for their helpful discussions.
This work was supported by the National Natural Science Foundation of China under Grant No.~12175015, No.~12304171,
No.~12088101; the Ministry of Science and Technology under Grant No. 2022YFA1402701; the Beijing Institute of Technology Research Fund Program for Young Scholars; and the Key Laboratory of Multiscale Spin Physics at Beijing Normal University under Grant No.~SPIN2024N01 .
The authors acknowledge the support of Tianhe 2JK at the Beijing Computational Science Research Center(CSRC) and
the Super Computing Center of Beijing Normal University.
\end{acknowledgments}

  %\clearpage
 % \bibliographystyle{apsreve}
  \bibliography{ref.bib}
  
  \clearpage
 %\appendix
     \renewcommand{\theequation}{S\arabic{equation}}
    \setcounter{equation}{0}
    \renewcommand{\thefigure}{S\arabic{figure}}
    \setcounter{figure}{0}

\section{Supplemental Material}
\subsection{Scaling of EE for $O(n)$ coplanar ordered system}

Take the $SU(2)$ AFM Heisenberg model on the triangular lattice as an example. 
The low-lying levels in each of the two subsystems separated by smooth boundaries, $A$ and $B$, are described by an $O(n)$ rotor with the effective Hamiltonian 
\begin{equation}
	H=\frac{{\bm S}^2}{2\chi L^d},
\end{equation}
where ${\bm S}$ is the total spin of the system, $\chi L^d$ is the effective moment of inertia with $\chi$ the magnetic susceptibility. 
$\Delta_{\rm tow} \sim 1/\chi L^d$ is the energy scale of the tower of states. 

Suppose the system is coplanar ordered, then following \cite{Kulchytskyy}, the interaction between two subsystems can be described by a harmonic oscillator with the energy scale of Goldstone modes
$\Delta_G \sim c/L$ where $c$ is the spin-wave velocity.
The fluctuations in the relative angular momentum are given by the ground state fluctuations of the harmonic oscillator
\begin{equation}
    \langle S_A^2\rangle \sim \frac{\Delta_G}{\Delta_{\rm tow}}.
\end{equation}
%where $\Delta_G\sim c/L$ is the Goldstone mode gap which is the scale of the lowest energy spin waves.
This fluctuation effectively cuts off subsystem rotor states that are accessed in the ground state at order $S_{\rm cut}^2\sim \Delta_G/\Delta_{\rm tow}$.  
We can estimate the total number of states below the energy scale $\Delta_G$ by integrating the degeneracy up to the cutoff
\begin{equation}
    %\Omega_A = \int_0^{S_{\rm cut}}  S^{2(N_G-1)} dS,
    \Omega_A = \int_0^{S_{\rm cut}}  S^{(n-2)2} dS,
\end{equation}
where the degeneracy of each level is of order $S^{(2n-4)}$, which is related to the number of Goldstone modes by  $S^{N_G-1}$ 
for the coplanar magnetic order\cite{Bernu_triangularHeis,Azaria}.
This gives the logarithmic correction to the area law $S_{\rm tow} \sim \ln \Omega_A$, which leads to 
Eq.(\ref{fssS2}) (and Eq.(\ref{newS2})) % into $b=(2n-3)/2$.
with number of Goldstone modes $N_G = 2n-3$ for the coplanar ordered $O(n)$ AFM. 
The formula obtained in \cite{Rademaker} is a special case of this result with $n=3$.

\subsection{ Polynomial fit of $\chi_\perp(L)$ and $\rho_s(L)$}
\label{fssrhos}

We analyze $\chi_{\perp}(L)$ and $\rho_{s}(L)$ to obtain the thermodynamic limit value $\chi_\perp$ and $\rho_{s}$, using the following expansions 
\begin{equation}
    \chi_{\perp}(L)=\chi_{\perp}(1+\frac{a_{1}}{L}+\frac{a_{2}}{L^2}+\cdots),
    \label{chi_fss}
\end{equation}
and
\begin{equation}
  \rho_{s}(L)=\rho_{s}(1+\frac{b_{1}}{L}+\frac{b_{2}}{L^2}+\cdots),
    \label{rhos_fss}
\end{equation}
where $a_i$ and $b_i$ are constants, and $\rho_{s}$ and $\chi_\perp$ is the spin stiffness
and the susceptibility at the thermodynamic limit $L \to \infty$, respectively.

Fitting Eq. (\ref{chi_fss}) and Eq. (\ref{rhos_fss}) up to the second order of $1/L$ to our QMC data,
 in the process, we gradually increase the smallest system size $L_{\rm min}$ in the analysis and achieve stable fitting for $L\geq 40$. The details of fitting results are listed in Tab.{\ref{tablechiq1}} and \ref{tablerho1}.
We see $\rho_s(L)$ and $\chi_\perp(L)$ converge to finite values. The final estimates 
of $\chi_\perp$ and $\rho_s$ are then obtained as $\chi_\perp=0.00288(6)$ and $\rho_s=0.135(3)$.

\begin{table}[thb]
	\caption{ Polynomial fit $\chi_{\perp}(L)=\chi_\perp+a_1/L+a_2/L^2$ to the data of $\chi_{\perp}(L)$. The range of the system sizes is from $L=8$ to $L=64$.}
    \begin{tabular}{c|c|c|c|c}
    \hline
     \hline
         $L_{\rm min}$ &  $\chi_\perp$    & $a_1$& $a_2$&$\chi_r^2$/P-value \\   
    \hline
                28 &  0.00275(3)  & 0.071(2)  &   0.41(3)                & 1.95/0.06   \\
    \hline
                32& 0.00280(4)   & 0.066(3)   & 0.52(6)                  &1.43/0.20\\
   \hline
               36&  0.00288(6)   & 0.058(5)   &  0.71(11)                & 0.71/0.61 \\
     \hline
               40&  0.00293(9)   & 0.053(8)   &  0.84(20)                & 0.75/0.56\\
 \hline
        \hline
     \end{tabular}
     \label{tablechiq1}
 \end{table}

\begin{table}[thb]
	\caption{ Polynomial fit $\rho_{s}(L)=\rho_s+b_1/L+b_2/L^2$ to the data of $\rho_s(L)$. The range of the system sizes is from $L=8$ to $L=64$.}
    \begin{tabular}{c|c|c|c|c}
    \hline
     \hline
         $L_{\rm min}$ &  $\rho_s$    & $b_1$   & $b_2$&$\chi_r^2$/P-value \\   
    \hline
                24     &  0.1286(7)  & 2.26(5)  &   3(1)                & 1.42/0.18   \\
    \hline
                28     & 0.130(1)    & 2.15(8)  &   5(2)                  &1.17/0.32\\
   \hline
               32      &  0.131(2)   & 2.0(2)   &  8(3)                & 1.07/0.38 \\
     \hline
               36      &  0.135(3)   & 1.7(3)   &  16(6)                & 0.51/0.77\\
 \hline
        \hline
     \end{tabular}
     \label{tablerho1}
 \end{table}

\clearpage

%\end{CJK*}  
\end{document}